# CERN antiproton target: Hydrocode analysis of its core material dynamic response under proton beam impact

Claudio Torregrosa Martin,[1,2,*] Antonio Perillo-Marcone,[1]
Marco Calviani,[1] and José-Luis Muñoz-Cobo[2]

[1]*CERN, 1211 Geneva 23, Switzerland*
[2]*Universidad Politécnica de Valencia, Camino de Vera s/n, 46022 Valencia, Spain*

Antiprotons are produced at CERN by colliding a 26 GeV/c proton beam with a fixed target made of a 3 mm diameter, 55 mm length iridium core. The inherent characteristics of antiproton production involve extremely high energy depositions inside the target when impacted by each primary proton beam, making it one of the most dynamically demanding among high energy solid targets in the world, with a rise temperature above 2000 °C after each pulse impact and successive dynamic pressure waves of the order of GPa's. An optimized redesign of the current target is foreseen for the next 20 years of operation. As a first step in the design procedure, this numerical study delves into the fundamental phenomena present in the target material core under proton pulse impact and subsequent pressure wave propagation by the use of hydrocodes. Three major phenomena have been identified, (i) the dominance of a high frequency radial wave which produces destructive compressive-to-tensile pressure response (ii) The existence of end-of-pulse tensile waves and its relevance on the overall response (iii) A reduction of 44% in tensile pressure could be obtained by the use of a high density tantalum cladding.

DOI: 10.1103/PhysRevAccelBeams.19.073402

## I. INTRODUCTION

The AD-Target system is the main particle production element of the CERN Antiproton Decelerator. Antiprotons are produced by colliding a proton beam of 26 GeV/c from CERN Proton Synchrotron (PS) with a fixed target made of a dense and high-Z material. The high energy collision of the proton beam with the atoms of the target creates a shower of secondary particles, and among them, antiprotons. The generated antiprotons travel through the target assembly and are collected downstream by a magnetic focusing device using their charge, after which, they are selected and magnetically conducted via the injection line to the antiproton decelerator (AD) complex for antimatter research experiments. Antiparticles should be identical to matter particles except for the sign of their electric charge. It is well known, however, that matter and antimatter are not exact opposites; nature seems to have a one-part in 10 billion preference for matter over antimatter. Understanding matter-antimatter asymmetry is one of the greatest challenges in physics today.

The relevance of the antiproton target is not only motivated by the significance of its purpose—to produce new particles to serve as a window to new physics—but for the challenge that involves to overcome the engineering problem of its design. The characteristics of antiproton production require a very compact target in order to avoid antiproton reabsorption in the surrounding material and to be as close as possible to a punctual source for the antiproton downstream collector system [1,2]. For this reason, a very thin rod of a high density material and a very focused primary proton beam have to be used. This results in extremely high energy depositions reached inside the target core as a consequence of each proton beam impact, which makes the antiproton target a unique case even among high energy particle-producing primary targets. As a matter of fact, calculations show that in each proton pulse impact the peak temperature rises up to 2000°C. Dynamic pressure waves of several GPa and strain rates well above $10^4 s^{-1}$ are easily exceeded as a consequence of the sudden expansion of the material. These waves propagate through the target and surrounding containment, where expected stresses are also considerably high [3]. All these phenomena turn into a real challenge to predict the target material response and behavior during operation. In addition, radiation damage including material embrittlement, swelling and noble gas production inside the target material may also play an additional role [4–6]. These damaging processes, however, are not considered in this publication and are subjects of ongoing studies.

After 2016 a new upgrade of the CERN AD facility is planned, including an optimized redesign of the full antiproton production system in order to enhance its performance, reliability as well as to guarantee the continuity of the antiproton physics for the next 20 years [7–9]. The present configuration of the CERN antiproton target dates back to the late 1980s and it was obtained after more than 10 years of iterations and improvements of the material choice and conceptual designs. Copper, tungsten, rhenium and iridium were employed during this decade. The current design consists of a core target in form of a 55 mm long, 3 mm diameter iridium rod—one of the densest elements on earth, second only to osmium—encapsulated in a 15 mm diameter graphite matrix; the whole assembly is in turn embedded into a water-cooled Ti-6Al-4V body as shown in Fig. 1.

[*]claudio.torregrosa@cern.ch

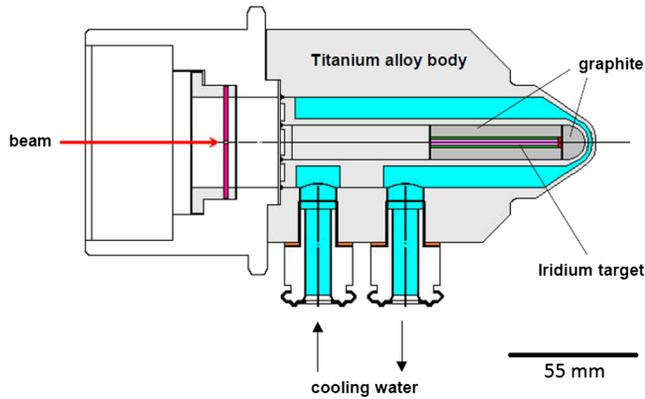

FIG. 1. Schematics of the water cooled antiproton target design used from 1987 to present day.

The only other antiproton production target existing in the world was the Pbar source at Fermilab, which was in operation from 1985 to 2011, proving antiprotons for p-pbar collisions at the Tevatron [10,11]. The conceptual design of this target was significantly different from the CERN one, consisting of a stack of disk of different materials which vertical axis was perpendicular to the primary proton beam. By raising and lowering the target discs it was possible to change the target material and by moving the target horizontally it was possible to change the material thickness exposed to the beam. In addition, the target was rotated between beam pulses to spread depletion and damage uniformly around the circumference. This concept, however, is not compatible with the CERN AD-target area since it requires us to have the targets inside a vault. Different materials were tested as well during the 26 years of operation of the Pbar source. Initially tungsten or tungsten alloy such as tungsten-rhenium. Later on, experiences with severe damage to the tungsten targets prompted a switch to copper, nickel or INCONEL®. Two main problems were reported as the cause of targets failure during operation. First, the rapid heating and expansion of the target material and successive shock waves causing mechanical damage to the target. Second, a molten channel formed in the target, in particular when using copper, nickel or INCONEL® targets [12], which have significantly lower melting temperatures.

The study of the dynamic response of the target material is fundamental in order to reduce the uncertainties concerning the target core structural state and to provide a robust design. Any damage and effective loss of density of the target material will lead to target exchange since the proton-target interactions and antiproton production will be reduced to the point that the target has to be replaced periodically to maintain the desired performance [13,14]. Unfortunately, post examination of failed targets is very complex due to high activation of the irradiated material. For this reason, a reliable methodology based on advanced computational tools and experimental tests is proposed in order to reduce uncertainties on the target response and to assess the choice of future design and candidate material. In the present study, the numerical part of this approach is shown with the aim of identifying the AD-target material working conditions and analyze its governing phenomena.

The thermomechanical effects induced in materials by high energy particle beam impact have been extensively studied in accelerator technology in a wide number of facilities around the world for the design of BIDs (beam intercepting devices) such as targets, collimators, and dumps. References [15–18] are a few among many examples. In these studies, the energy deposition map arising as a consequence of proton beam interaction with matter is calculated by means of Monte Carlo simulations and then applied as an internal energy load to thermomechanical finite element (FE) solvers. The thermomechanical effects induced in solid were also studied analytically in [19], where dynamic stresses were solved for a single circular traversal section assuming impact of a Gaussian beam. All these calculations, however, are normally performed under assumptions of material elastic regime and considering quasistatic analysis, focused toward steady state conditions since raise of temperatures are within the order of hundreds of degrees per pulse. Unfortunately, these assumptions are by far out of reality for our problem of study, where deposited energy is much higher and faster. The material will certainly deform plastically or break. As a matter of fact, only the previously described target of the Pbar Source at Fermilab has worked in a similar range of energy deposition density as CERNs AD-target.

Hydrocodes are a family of advanced highly nonlinear computational tools which are able to fully simulate the dynamic response of materials [20]. These codes numerically solve the equations arising from mass, momentum and energy conservation. They are able to simulate the material response at different thermodynamic and mechanical states (including response beyond plasticity and fracture) taking into account strain rate and temperature dependence by coupling these equations with material models as equations of state (EOS), strength, and failure models. Hydrocodes have been historically developed and used in ballistic and military applications where the dynamic load originates from high velocity impacts and explosions. However, they have started being applied in accelerator technology for simulations of BIDs, where the load is generated by the sudden thermal expansion of the material impacted by particle pulses and the subsequent propagation of the produced stress waves. Current programs for development of accelerator technologies require higher and higher particle beam energies and intensities, which push engineers to use these codes to investigate mechanical damage in materials hit by high intensity beams as a consequence of abnormal operation, in which hydrodynamic behavior of materials is expected. Simple 2-D hydrodynamic calculations were already used at CERN in 1984 to assess the response of solids and liquid materials to the impact of proton pulses [21]. In a similar way, hydrodynamic calculations for assessing the tolerance of the SSC beam dump to a hypothetical full impact of 20-TeV proton beam was done in 1993 [22]. More recent and comprehensive analyses using AUTODYN ® were done for the study of the structural behaviour of the Main LHC absorber block in case of a total beam dilution failure [23]. Numerical studies of high dynamic transient effects of pulse beams on BIDs can be found in [24,25], where hydrodynamical calculations for uranium beam impacted on copper targets for ISOLDE were performed. Very recent publications for research on new collimator materials using ANSYS AUTODYN® can be found in [26,27], while using as well LS-DYNA® in Refs. [28–31]. In the same way, literature including hydrodynamic calculations using BIG2 code for the design of solids targets for the FAIR facility in Germany can be found in [32,33] as well as for studying the penetration of LHC proton beams in carbon targets and hydrodynamic tunneling effect [34].

The goal of the present paper is therefore, the application of hydrocodes in order to simulate the dynamic response of the AD-Target core after one proton pulse impact, in order to gain important information concerning its response, understand the dominating phenomena and learn lessons which could define strategies to provide a new design which should be as robust as possible.

## II. COMPUTATIONAL APPROACH

Figure 2 shows a scheme of the methodology employed in this study. First, an energy deposition map in the target material as a consequence of the proton beam-target atoms interaction is calculated by means of FLUKA Monte Carlo simulations, a highly advanced particle transport code developed at CERN which is able to simulate the energy released by these nuclear reactions also taking into account secondary reactions of the particle cascade generated by the proton beam when going through the target material [35]. This energy deposition map is then applied as an internal energy load to the commercial hydrocode ANSYS AUTODYN® via an internal user subroutine and the user function `EXSIE3`. The energy is applied consistently with the PS beam parameters, whose proton beam is composed by 4 proton bunches spaced by 105 ns, bunch length of 30 ns for a total pulse intensity of $1.5 \times 10^{13}$ protons per pulse [9]. The total energy deposited in the target core is approximately 1.34 kJ, which results in 11.17 GW considering that it is deposited only in 120 ns. Moreover, the power is applied in a small volume of the target core, producing a mean power density of $2.87 \times 10^4$ TW/m$^3$.

It is important to note that the interaction between FLUKA and ANSYS AUTODYN® is only a one-way coupling, which means that the reduction in the deposited energy during the pulse impact as a consequence of the change of target material density is not taken into account. This is considered a valid assumption by the authors as the maximum decrease of density during the 430 ns pulse burst due to the increase of temperature is in the order of 2% of the initial density, since the material is not melting.

ANSYS AUTODYN® provides a good variety of solvers for the numerical resolution of the mass, momentum and energy equations, enabling it to obtain the full dynamic response of the material. For the present study, a Lagrangian solver is used as the deformation of the material is produced by its thermal expansion and no important mesh distortions occur.

### A. Material models employed by the hydrocodes

There are three main ingredients necessary to perform hydrocode calculations; equation of state (EOS), strength model and failure model. These are the material laws which, coupled with the mass, momentum and energy equations, define the material behavior. Full confidence on simulations relies on the use of proper material models for the regime of application [20]. Unfortunately, the material constants associated to many of these models are rarely available in literature, and are often classified as it is drawn from military research. Moreover, most of the AD-target material candidates such as iridium or tungsten-rhenium alloys are uncommon and they have never been studied at any dynamic regime at all, even less so for the ones present in the case of study, with temperatures above 2000°C and strain rates above $10^4 s^{-1}$. It is for this reason that the calculations here are done assuming pure tungsten as a target material, for which more material models are available. Tungsten-rhenium alloys would probably present a more ductile response than pure tungsten, however several studies had shown that the initial more ductile behavior of tungsten-rhenium in comparison to pure tungsten rapidly vanishes when exposed to radiation due to the formation of precipitates at the grain boundaries [36–38]. In addition, tungsten was and is a strong candidate material and the lessons learned from this study depend mainly on the pure dynamic response, making it possible to extrapolate theses conclusions to other candidate materials. Furthermore, experimental tests under real proton beam and similar conditions as reached in the AD-target are also foreseen to validate these phenomena.

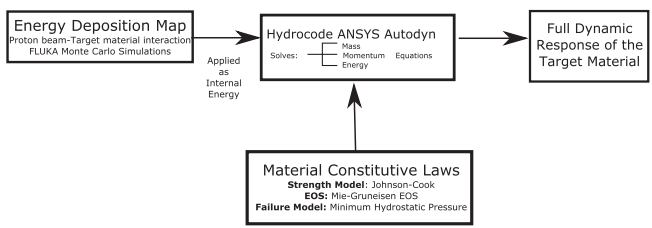

FIG. 2. Scheme of the methodology employed in this study, including the two computational tools FLUKA and ANSYS AUTODYN®.

### 1. Equation of state (EOS)

An equation of state is a model which expresses the relation between thermodynamic state variables as pressure, energy and density. EOS were originally developed for ideal gases and subsequently extended to all states of matter [39]. In continuum mechanics, the stress tensor is usually decomposed in a hydrostatic component and deviatoric component. The EOS governs hydrostatic component associated with the change of pressure as a function of volume and internal energy and it is strongly correlated with the wave propagation velocity and shock response. Ideally, the EOS includes also change of phase of the material, although in practice it is difficult to include them in a single formulation. There are different EOS formulations depending on the regime of applications and assumptions, Linear EOS, Mie-Grüneisen, Puff, Tillotson EOS [40] and SESAME tables [41] are some of them. For this study Mie-Grüneisen is employed. This model is one of the most widely used EOS and one of which more parameters are available in the literature, working especially well in the regime of not very high pressures where the relation between shock speed and particle speed can still be considered linear. The AUTODYN® implementation of Mie-Grüneisen takes as well into account differences between compression and tensile states. This formulation does not take into account change of phase, which is perfectly valid for the case under study—where melting of material does not occur due to the high melting point of the candidate materials—and the pressures reached are in the order of GPa's, which in terms of shock physics, are relatively low.

### 2. Strength model

The strength model, on the other hand, defines the deviatoric part of the stress-strain tensor, i.e., it governs the changes in shape as well as deformation beyond plasticity (flow stress). The strength model is therefore a constitutive relation that links stress with strain for different strain rates and temperatures. There are a wide number of strength model formulations depending on strain rate-temperature influence assumptions, type of material, and regime of application. Some of them are purely empirical as Johnson-Cook (J-C) [42], which is one of the simplest and more widely used strength model, or more physical based as Zerillini-Amstrong [43] which is based on dislocation dynamics. Other formulations are Steinberg-Guinan [44], Steinberg-Lund [45] or the mechanical threshold model [46]. For the present study, the J-C strength model was used, which takes into account the temperature and strain rate dependence on the material as well as its response beyond plasticity. This model is relatively simple, since it considers the influence of temperature and strain rate in an uncoupled way, but has been shown to provide good agreement between experiments and simulations in other studies of proton beam impact phenomena on materials [26]. The parameters employed for the strength model of pure tungsten were obtained by Hopkinson bar tests [47,48] at different high temperatures and high strain rates by K. T Ramesh et al. [49] and are partially extrapolated by the code to the regime reached in the AD-target.

### 3. Failure model

Finally, the third necessary element to simulate realistic material behavior is a failure model. A material cannot withstand local tensile stresses greater than its limits. When this threshold is exceeded, the code assumes that the material can no longer sustain any shear stress or any tensile pressures in the corresponding element. There are two main types of failure models used by the hydrocodes; (i) the material will instantaneously fail when it locally overcomes a limit for one or a certain number of variables (such as strain to fracture, tensile hydrostatic stress, maximum principal stress) and (ii) The failure model is based on the cumulative damage of the material when a certain variable is reiteratively exceeded [50]. The first one is generally associated to brittle failure mode while the second with ductile fracture [43]. In the present study, a model corresponding to the first type is employed, named the minimum hydrostatic pressure model [51]. This model assumes that the material fails when a certain negative pressure is reached. The arguments supporting the use of this model are the large tensile pressures reached in the target and the known brittleness of the material candidates. For this study a spall strength of $-2.6$ GPa was considered, estimated by J. Wang et al. using laser-induced stress waves on polycrystalline tungsten [52].

### B. Computational domain

Simulations are performed using the reference geometry shown in Fig. 3(a), which corresponds to the target core and containing graphite matrix. In addition to this configuration, two other configurations have been investigated, including a copper and tantalum cladding up to 1 mm thickness around the target core, [Figs. 3(b) and (c)]. The aim of these simulations was to investigate the influence of the impedance mismatch between target core and surrounding materials on the pressure wave response inside the target. The post processing tool of AUTODYN® allows us to add gauges in specific points of the geometry in order to check the time history of a selected variable. The results presented in the plots of Figs. 5–9 correspond to the position of the gauges shown in Fig. 3. For a deep understanding of the shock wave phenomena and interpretation of the results, the simulations were performed in a step by step way, increasing gradually the level of complexity of the models, i.e., from a single material (only the target core) to multiple materials (graphite matrix and cladding), as well as from assumptions of perfectly elastic material response, up to implementation of plastic models and finally, failure models.

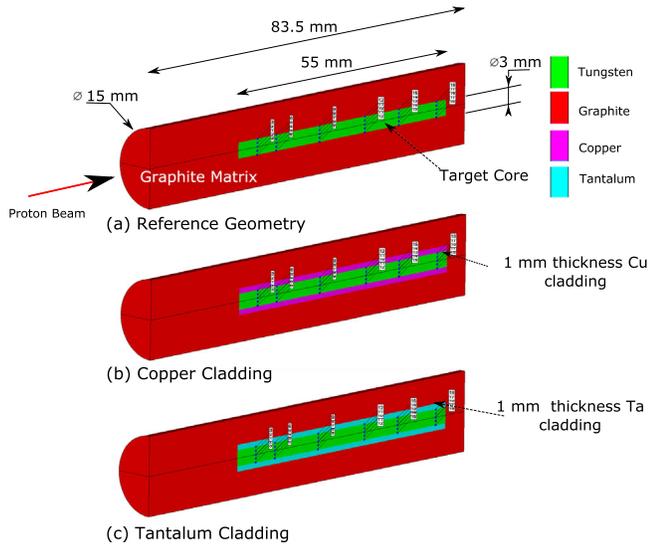

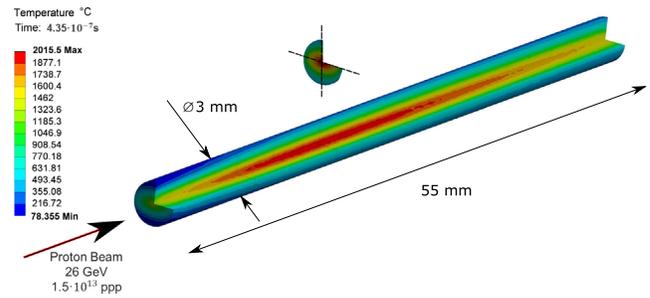

FIG. 4. 3/4 cut view of the temperature profile in the target core at the end of a single 26 GeV proton beam impact.

FIG. 3. Half-view geometry of the different configurations studied in the present work: (a) W target surrounded by graphite matrix (b) W target with Cu cladding surrounded by graphite matrix (c) W target with Ta cladding surrounded by graphite matrix.

From the point of view of physical constraints, the model is considering that the external graphite envelope can freely expand in every dimension. This assumption does not affect at all the dynamic response of the center core in the time window of interest, as the phenomena of study are that fast that there is not enough time for the speed of sound to travel and come back from the periphery of the computational domain (the external graphite envelope).

### III. RESULTS AND DISCUSSION

#### A. Adiabatic increase of temperature in the target

Figure 4 shows the temperature profile in a 3/4 cut view of the target core at the end of a single proton pulse impact, i.e., at 430 ns. This simulation assumes that the rod was at room temperature prior to the pulse impact. The profile shows the increase of temperature generated as a consequence of the proton-target material interaction, which reaches 2000 °C at the center of the rod. In addition, a temperature gradient of up to 1800 °C in only 1.5 mm of the radial direction takes place. The expansion of the material associated to this sudden rise in temperature causes the successive dynamic waves studied in detail in the next sections.

#### B. Elastic and plastic response of the target core

##### 1. Simulations assuming perfectly elastic material

Figures 5 and 6 show respectively the short and long transient pressure response in the center of the target core (gauge 7) and periphery (gauge 9) assuming a perfect elastic material. This assumption is by far out of reality as tungsten's yield strength is no more than 800 MPa, well exceeded in this study. Therefore, the pressure values shown in these plots can only be taken into account as a qualitative way since the energy diffused by plastic deformation processes is not taken into account. Simulations assuming elastic material, however, allow a better understanding of the pressure wave propagation as no damping of the waves on the material occurs.

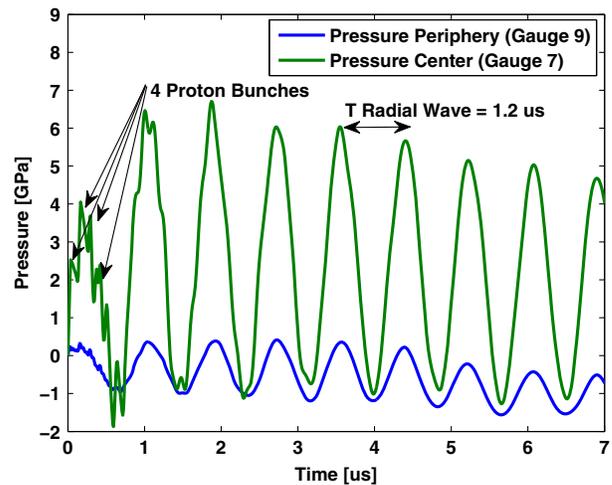

FIG. 5. Pressure response at the center (gauge 7) and periphery (gauge 9) during the first 7 $\mu$s after the 430 ns -four bunches- proton pulse impact under infinitely elastic material assumption (reached pressures only for qualitative information). The plot clearly shows the existence of a radial wave with a 0.85 $\mu$s period. Note that the oscillation takes place in phase in the center and periphery, meaning that the rod material simultaneously expands and contracts within its transversal section.

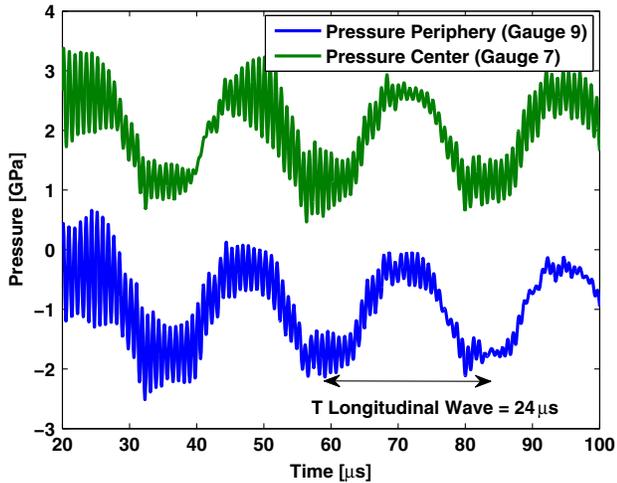

FIG. 6. Pressure response at the center (gauge 7) and periphery (gauge 9) during 20 μs–100 μs under infinitely elastic material assumption (reached pressures only for qualitative information). The plot clearly shows the presence of high frequency radial and low frequency −24 μs period- longitudinal waves.

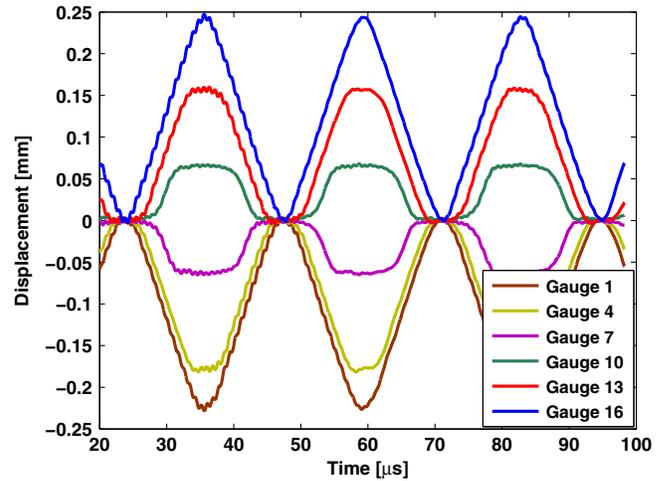

FIG. 7. Plot of the axial displacement at different longitudinal positions of the target core after proton pulse impact and under infinitely elastic material assumption. The plot clearly shows how the material of the target core longitudinally expands and contracts with a period of 24 μs.

Figure 5 shows the pressure response in the target during the first 7 μs. A four stepped rise of pressure corresponding to the four impacts of four proton bunches spaced by 120 ns can be observed. Then, a clear high frequency pressure wave response takes place. This wave has a period of 0.85 μs, corresponding to the time needed for the induced expansion to travel and come back from the rod circumferential surface. The pressure wave in the center and periphery (gauges 7 and 9) are practically in phase, meaning that the whole section of the rod expands simultaneously. This is due to the fact that the sudden rise of temperature in the material not only happens at the center but also at the periphery of the rod, where the increase of temperature is also significant (above 300 °C, as can be seen in Fig. 4). The consequence is that the pressure waves are not just traveling from a central, localized area, but there are infinitesimal pressure waves with infinitesimal departing points along the rod material. The amplitudes of these pressure waves are attenuated as they travel to the peripheral surface, and amplified as they do it through the center due to the cylindrical geometry. The pressure, and other variables, oscillations are the result of the interactions of all these waves at each single point.

Figure 6 plots the response of the transient analysis in an expanded time window, from 20 μs to 100 μs, allowing us to appreciate, in addition to radial high frequency wave, a longitudinal pressure wave. This wave has a period of 24 μs, corresponding to the resultant period consequent of the interaction of the infinitesimal waves traveling along the 55 mm longitudinal distance. It is important to note as well that this longitudinal pressure wave is in phase at all the longitudinal points of the rod. This can be easily appreciated in Fig. 7, where the longitudinal displacement at different longitudinal points of the rod axis (gauges 1,4,7,10,13,16 in Fig. 3) is shown. The coordinate reference system in the model is placed at longitudinal center of the rod, higher displacements are symmetrically reached at gauges 1 and 16, which are close to each of the rod ends, while displacement is reduced when symmetrically approaching the longitudinal center of the rod—gauges 7 and 10. The whole rod is therefore longitudinally expanding in a simultaneous way within all its volume, in a similar manner as it was happening radially. The reason for this phenomenon, again, is that the temperature rise caused by the deposited energy is also covering a great part of the longitudinal volume of the core rod. An important point to be detailed here is that the velocity of propagation of the radial and the longitudinal waves do not match. This can be easily found out by just dividing the longitudinal and radial period by its corresponding radial and longitudinal distances. Initially one may think that the nature of the origin of the wave is the triaxial expansion of the inner material and therefore the wave should travel through the bulk material at the same speed, no matter the direction. This however is not perfectly true, since ultimately, the triaxial expansion in a single point and its propagation is governed and constrained by the surrounding material as well. In that sense, and in this specific case where the geometry is that small, the amount of surrounding material does matter. In fact, the resulting radial oscillation has a faster speed of propagation than the longitudinal one since there is much less surrounding material limiting the expansion radially than longitudinally.

#### 2. Simulations including J-C strength model

Figure 8 plots the pressure response during the first 7 μs with a simulation applying J-C strength model, which takes into account plasticity. This simulation is therefore much more realistic than the previous one, which assumed an elastic model. Figure 8 shows how the pressure wave experiences a fast damping due to the energy loss associated to the plastic deformation processes, which provoke an attenuation of the wave in just 20 μs. This fast attenuation prevents the appreciation of the lower frequency longitudinal waves presented in Fig. 6. In addition, another important difference concerning the wave response with respect to the elastic assumption is the compressive-to-tensile nature of the wave, reaching values up to 4 GPa and −5 GPa in the center of the rod during the first oscillations. This compressive-to-tensile response takes place also at the periphery, but with significantly lower amplitudes. This relevant difference in pressure distribution compared to the elastic assumption—Figs. 5 and 6—where the oscillations of pressure at the center of the rod were taking place always in compressive states is due to the limitation of the deviatoric component in the stress tensor by the plastic limit of the material. The proper predictions of these high tensile states are fundamental, since they are well above the spall strength of the material and could cause internal cracking and fragmentation in the inner part of the target rod, producing a high loss of density which will lead to a significant decrease of antiproton yield. This fact also states the importance of using proper strength models beyond plasticity, since wrong assumptions will not only affect the magnitude of stress-strain response but can change completely its distribution along the geometry.

Another interesting phenomenon that deserves being explained in detail can be observed in Fig. 9, which shows the pressure response and its time derivative at the center of the target during the first 1.5 μs of the transient. This plot includes therefore the four sudden rises of pressure corresponding to the four impacting bunches. The interesting phenomenon occurs at the end of each proton bunch impact —105 ns, 210 ns, 315 and 420 ns—and consists of an abrupt decrease of pressure as a consequence of the inertia after the sudden expansion of the material, as can be easily seen when plotting the pressure derivative (green line in Fig. 9). This end-of-bunch decrease of pressure is significantly relevant since it may be the cause of the strong compressive-to-tensile oscillation response present in the center of target when it is in constructive interference with the natural radial wave as it is demonstrated in the next section. The plot of the pressure derivative in Fig. 9 allows us to observe as well four delayed waves corresponding to the reflection of the initial perturbation produced at the beginning of each single bunch impact. The interactions of these reflective waves are the responsible of the small oscillations on the radial fundamental wave observed during the first period on Figs. 5 and 8.

### C. Importance of the proton pulse length on the pressure response

The previous subsection allowed pointing out the existence of an end-of-bunch tensile wave taking place due to the inertia effects. We have investigated whether these tensile waves can have an important influence on the overall response. For this reason, a parametric analysis assuming proton impacts with different pulse length has been carried out. For this analysis it is assumed that all the energy is deposited in the target in continuous pulses (single bunch). The total energy applied for each of the scenarios is the same, with the only difference of the duration of the pulse. A J-C strength model including plastic deformation of the material is used. Figure 10 shows the pressure response in the center of the target assuming times of proton pulses from 0.1 $\mu$s (blue) to 1.2 $\mu$s (pink) in steps of 0.2 $\mu$s. As can be seen in the plot, the magnitude of pressures reached changes completely depending on the length of the pulse, ranging from $+/-9$ GPa with 0.1 $\mu$s pulse to $+1$ GPa only in compressive stress with 1 $\mu$s. The reason for this change is not only the evident fact that the shorter the pulse the greater the power and inertia but the influence of when the pulse is finishing. It can be observed in the plot that a tensile wave which changes the trend of the pressure wave always takes place at the end of the pulse. The effect of this end-of-pulse tensile wave differs significantly depending on whether the material is going to tensile or compressive states. When the material is going to tensile states, the end-of-pulse wave and radial wave produce a constructive interference leading to very large tensile stresses in the first period of the radial wave, which propagates to the rest of the transient pressure response. This can be easily appreciated in Fig. 11; higher end-of-pulses pressure drops take place in the cases where pulses finish when $DP/Dt < 0$. This is the case of pulses of 0.43 $\mu$s length (the real of the AD-target), 0.6 $\mu$s length and 1.2 $\mu$s length. On the other hand, in the pulses which end between 1/2 and 3/4 of the period of the radial wave (when the pressure is going to compressive states and therefore $DP/Dt > 0$) the end-of-pulse tensile wave is counteracted by the radial wave. This is the case of 0.8 $\mu$s length and 1 $\mu$s pulses in Figs. 10 and 11, where the tensile pressure reached is below $-1$ GPa. This phenomenon is even clearer when comparing the 1 $\mu$s length pulse (yellow) with the one of 1.2 $\mu$s. Higher tensile and compressive stresses are reached in the latter, even if the energy is deposited slower in this case. The only exception to this response takes place with the pulse 0.1 $\mu$s, where due to the high inertia, the amplification of the end-of-pulse tensile wave takes place even if $DP/Dt > 0$. In any case, a straightforward conclusion that can be drawn from this analysis is that a way to reduce the harmful tensile stresses reached in the AD-target, given that the length of the pulses is strongly constrained by the PS and AD operation, would be to increase the diameter of the target core to 5–6 mm in order to increase the period of the radial wave and avoid its constructive interference with the end-of-pulse target. This diameter increase could however increment the antiproton reabsorption at the periphery of the target, in particular of the antiprotons created in the interactions at the beam halo. This effect may not be very relevant with appropriate focusing but it needs to be demonstrated via FLUKA and more detailed trade-off analyses.

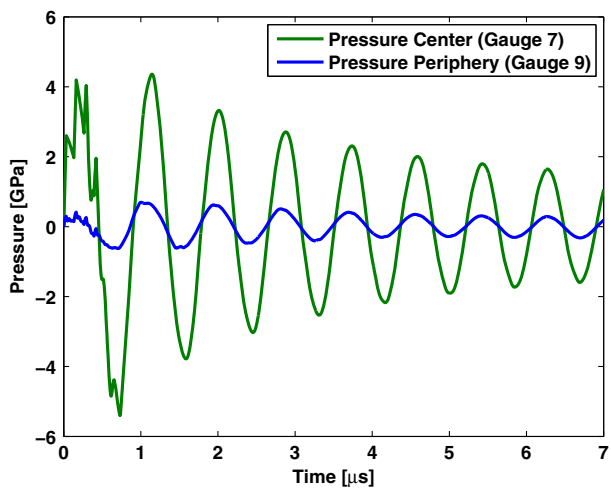

FIG. 8. Pressure response at the center (gauge 7) and periphery (gauge 9) of the target core after a 430 ns proton pulse impact. This simulation considers J-C strength model, which takes into account the material response beyond plasticity. Note the change in the pressure distribution in comparison to the pure elastic material assumption (Fig. 5).

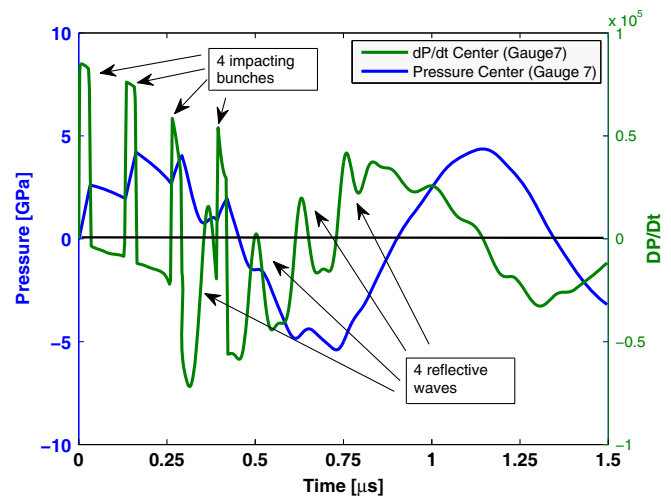

FIG. 9. Plot of the pressure response (blue) and its time derivative (green) in the center of the target core (gauge 7) during the first 1.5 μs after proton pulse impact. The plot clearly shows the discontinuities in the pressure derivative corresponding to the beginning and end of each of the four proton bunch impacts, the produced tensile wave at the end of the bursts, and four delayed waves, corresponding to the reflection of the four bunch impacts. This simulation is considering a J-C strength model for the tungsten target core.

### D. Influence of cladding and application of failure models to the cases of study

Another studied strategy in order to decrease the magnitude of the tensile pressure in the center of the target core could be the addition of an external cladding. It is well known in the shock and explosive physics literature that the use of multilayer armatures can effectively reduce the shock-induced tensile stresses in explosively expanding materials. The key point resides in the acoustic impedance mismatch between the different materials in a multilayered media, which governs the fraction of the pressure wave energy that is transited/reflected between them [53,54]. The greater the impedance mismatch, the greater the portion of the shock energy reflected at the boundary, and the smaller, the greater portion transmitted to the adjacent media. The ideal situation from the shock wave view point would be that no impedance mismatching exists between the target core and surrounding material, therefore all the energy of the radial pressure wave is transmitted to the adjacent media instead of being reflected to the core as a tensile and destructive wave. This possibility is studied numerically in the present section by assuming three different scenarios: (i) No cladding between the 3 mm diameter high density target core and the surrounding graphite matrix [Fig. 3(a)]. This is the case of the current design of the target, where impedance mismatch is very big due to the large difference between tungsten and graphite's density and sound velocity ($\approx 30$ times ratio in acoustic impedance between target core and graphite). (ii) Copper cladding of 1 mm thickness [Fig. 3(b)]. ($\approx 2$ times ratio in acoustic impedance between target core and copper). (iii) Tantalum cladding of 1 mm thickness [Fig. 3(c)]. ($\approx 1.4$ times ratio in acoustic impedance between target core and tantalum). For these calculations, strength models which take into account the response beyond plasticity are considered as well for the cladding materials as shown in Table I. No failure model is considered in order to see the propagation of the wave. The contact boundary between the target core and surrounding materials is assumed to be perfect by the code, i.e., the mesh is considering a continuous body with shared nodes at the interface, so no contact algorithm is necessary. This is done in order to avoid sensitivity of the mesh and the nonlinearity of the contact algorithm parameters for the comparison between results, reducing the problem to a pure dynamic phenomenon. In reality a kind of contact like this could be achieved by bounding via hipping, a possibility that is currently under investigation.

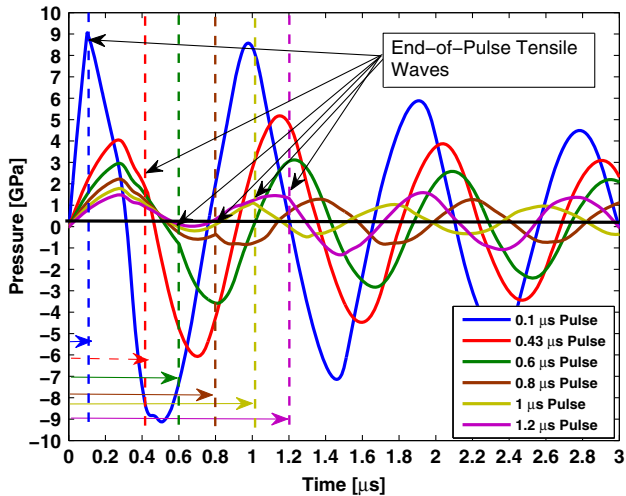

FIG. 10. Parametric study showing the pressure response in the center of the target core (gauge 7) for different proton pulse lengths. End-of-pulses tensile waves can be easily appreciated in the plot, as well as their amplifying effect when they are in phase with the natural radial waves. Compare 0.8 $\mu$s pulse length (brown) with 1.2 $\mu$s (purple), higher tensile pressure are reached in the former even if energy is deposited in a slower fashion.

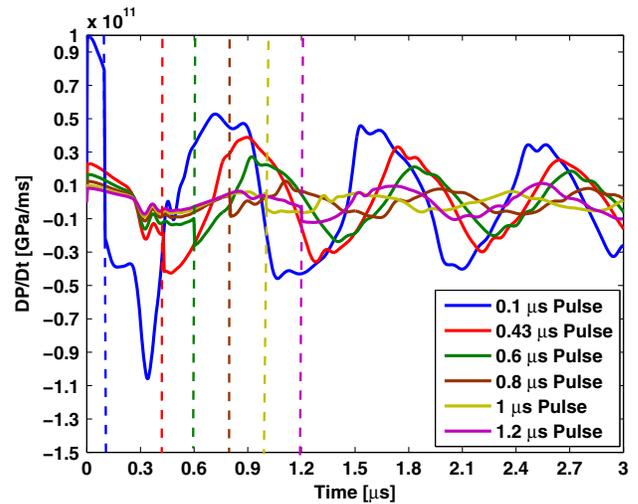

FIG. 11. Same analysis as in Fig. 10 but showing the pressure derivative for different pulse lengths. The influence of the end-of-pulse can be seen in the drop of the pressure derivative at the end of the pulses. It can be easily seen as well that, except on case of 0.1 $\mu$s length, these discontinuities in pressure derivative are more accentuated when the wave is going to tensile states, i.e. $DP/Dt < 0$.

TABLE I. Material models employed by the simulations.

| Material | EOS | Strength Model | Failure Model |
|---|---|---|---|
| Tungsten | Mie-Grüneisen | Johnson Cook[49] | Minimum Hydrostatic Pressure $P_{min} = -2.6$ GPa [52] |
| Graphite | Puff | Viscoelastic [55] | Principal Tensile Failure Strain |
| Copper | Mie-Grüneisen | J-C | … |
| Tantalum | Mie-Grüneisen | Von Mises (Yield strength 500 MPa) | … |

Figure 12 shows the pressure response at the center of the target core during the first 2 $\mu$s for the three scenarios of interest. As shown in the plot, the magnitude of the tensile pressure in the radial wave is significantly reduced when using copper and tantalum cladding. The reason for this reduction is probably a combination of 3 phenomena: first, the fact of the better impedance match between the core of tungsten and tantalum which causes a higher fraction of the energy of the pressure wave to be transmitted to the cladding instead of coming back as a tensile wave. Second, the shift in the period of the radial wave due to the larger diameter as a consequence of the cladding, which could partially avoid the constructive interference between the end-of-pulse wave and the radial ones explained in the previous section. Third, the plastic deformation in the tantalum or copper cladding material, which absorbs energy from the pressure wave, avoiding its reflection to the target core.

This reduction by 44% in the maximum tensile pressure reached in the center of the target when using tantalum can be very important from the point of view of target survival and increase of antiproton yield, especially if it is enough to reduce the tensile pressure below the spall strength of the material (which in tungsten was identified at 2.6 GPa). This can be seen in Fig. 13, where the minimum hydrostatic pressure failure model is applied in the three cases of interest after a single pulse of proton beam. As observed in the figure, for the case of a tungsten target without cladding, a large part of the volume of the target core is fragmented, leading to a reduction of antiproton production due to loss of effective density. The use of tantalum cladding, on the other hand, can efficiently reduce the volume of the fragmented region. The calculations show that only a $\approx 1\%$ volume is fragmented with the Ta cladding solution in comparison with the noncladded one. In this figure can be seen as well how most of the material of the target core experiences plastic deformation. One of the main challenges is therefore to understand what would be the real effect of this plastic deformation and its damaging influence. In addition, the impact of subsequent pulses could certainly increase the fragmented volume presented here, but in any case it is shown than Ta cladding solution is much more favorable than the noncladded one. Another open point, not covered in this study, is the effect of plastic deformation at the core-cladding interface, which could lead to cladding detaching and therefore reduce as well the improvement of the cladding solution obtained. This possibility will be studied experimentally in future works.

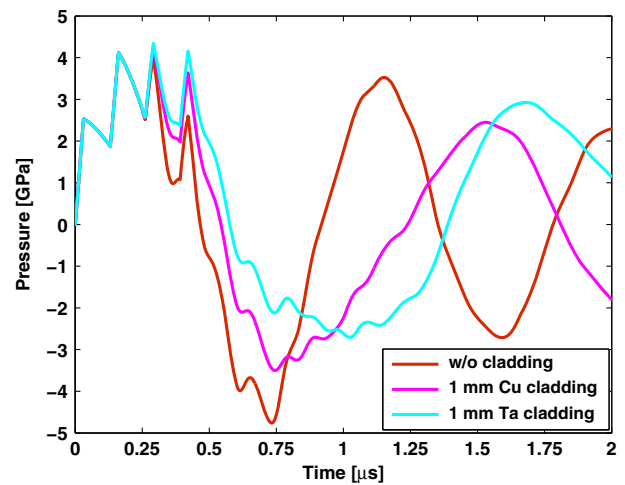

FIG. 12. Pressure reached in the center of the target core (gauge 7) for the three scenarios of study. Assuming Johnson-Cook strength model for the tungsten core and models considering plasticity in the cladding consistently with Table I. No failure models are considered. The plot shows the reduction of the reached tensile pressure when using Ta cladding.

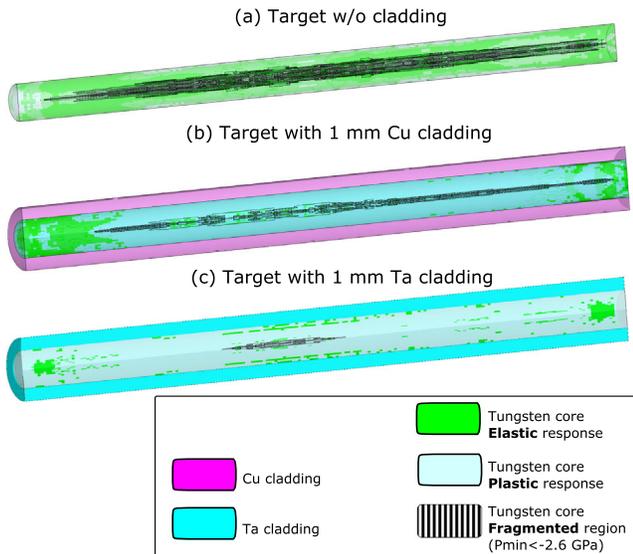

FIG. 13. Figure showing the state of the target core material after a proton pulse impact considering minimum hydrostatic failure model in tungsten. ($P_{min} = -2.6$ GPa). The figure shows results for the three scenarios of study, demonstrating the efficacy of Ta cladding for reducing the internal core damage.

## IV. CONCLUSIONS

In the present study a detailed analysis of CERN antiproton decelerator target has been carried out. This analysis includes a summary of the existing literature of the current design, which dates from the late 1980s, as well as of the previous designs built during the 10 years of iterations which led to it. This study emphasizes the extreme conditions reached in the 3 mm diameter, high density target core, in terms of adiabatic increase of temperature of more than 2000 °C each time the proton beam hits the target and the subsequent pressure wave and dynamic response. These facts probably make the AD-target the most dynamically demanding high energy target currently in operation around the world. The work does not cover the effect of further impact of consecutive pulses since their progressive damaging effect is subjected to the state of the material after the first pulse, propagating the already existing uncertainties in the calculations. In any case, it is quite obvious that further pulses will extend the level of target damage, although the generated pressure wave and its propagation through the material will be certainly different from the one presented here due to the internal fragmentation of the target.

The understanding of the dynamic phenomena induced on the target core as a consequence of proton beam impact turns out to be fundamental in order to provide a robust future design for the next 20–30 years of operation of the antiproton decelerator. Hydrocodes are a very powerful numerical tool, historically used for shock and impact physics, which are the only way to properly simulate this dynamic response of the target. The application of them to the target core is even more interesting, since its small and simple cylinder geometry allows us to easily appreciate the propagation of the generated pressure waves and understanding fundamental aspects about them. Simulations starting from simple models assuming perfectly elastic material, while adding gradually more complex models which take into account plasticity or fracture, helped as well for this purpose. In this way, three main phenomena have been identified and described in this study: (i) The identification of radial and longitudinal pressure waves present in the target core with a period of 0.85 $\mu$s and 24 $\mu$s respectively, the high frequency radial wave being the most important and critical one since the longitudinal one is rapidly damped by the plastic deformation of the material. The radial wave on the other hand produces huge oscillating compressive-to-tensile pressure response in the center and periphery of the target core, probably leading to its internal fracture in the first oscillations, when reaching tensile stresses above the spall strength of the material. (ii) The existence and importance of the end-of-pulse tensile waves, consequence of the inertia of the heated material expansion. These tensile waves become especially important when they are in phase with the tensile part of the compressive-to-tensile radial wave described above. This is the case of the AD-target core due to its small diameter of only 3 mm. Avoiding this constructive interference will be a way to reduce this limiting phenomenon. (iii) The possibility of using a 1-mm thick cladding surrounding the 3-mm diameter target core as a way to decrease the acoustic impedance mismatch between the target core and graphite containing matrix, with the aim of decreasing the tensile pressure reached in the center of the high density target core. Simulations show that a reduction of up to 44% in the CERN ANTIPROTON TARGET: HYDROCODE … PHYS. REV. ACCEL. BEAMS 19, 073402 (2016)maximum tensile pressure can be achieved by a tantalum cladding.

Nevertheless, one must be aware of the intrinsic limitations of these simulations, which are using material models based on simple assumptions and experimentally fitted as J-C can be. These models are applied here to a much more complex case than a simple mechanical impact test. The nature of the load itself, the variation of temperature and strain rates in this short time, will lead to lots of phenomena which cannot be considered by the code, like possible re-crystallization processes if there is enough time for it, presence of preexisting defects, material mechanical hysteresis or radiation damage. To these phenomena one has to add as well the effect on accuracy of the simulations due to the partial extrapolation of the used strength models, which had to be done because of the lack of experimental data in the literature at the extreme regime reached.

For this reason, the analysis in the current study only represents the numerical approach of the antiproton target redesign process. An experiment in the testing facility HiRadMat at CERN is foreseen to experimentally complete this numerical work. In this experiment different target materials as Ir, W, Mo, Ta, TZM and W cladded in Ta will be irradiated with high energy proton beam from the SPS, recreating the same conditions as reached in the real AD-target and exposed here. The goal of this experiment will be to validate these calculations, gain experimental insights of the real material response in these conditions and assess new potential candidate target materials for a future design.

## ACKNOWLEDGMENTS

The authors want to thank Herta Richter for the development of the subroutine used to import the energy deposition from FLUKA to AUTODYN®. In addition, the authors gratefully acknowledge the advices of Professor Lorenzo Peroni, which helped with the interpretation of some of the results, as well as the English proofreading of this document kindly done by Ayse Karatepe.[1] D. Möhl, Production of low-Energy Antiprotons, Hyperfine Interact. 109, 33 (1997).

[2] M. R. Bellone, G. del Torre, M. Ross, and P. Sievers, CERN Technical Report No. SPS/80-9/ABT, 1980.

[3] T. W. Eaton, CERN Technical Report No. PS/AA/Note 85-1, 1985.

[4] T. W. Eaton, CERN Technical Report No. PS/AA/Note 85-4, 1985.

[5] T. W. Eaton, CERN Technical Report No. PS/AA/Note 85-9, 1985.

[6] T. W. Eaton and C. Carter, CERN Technical Report No. PS/AA/Note 85-11, 1985.

[7] T. Eriksson, CERN Technical Report No. AB-2008-068 OP, 2008.

[8] T. Eriksson, M. E. Angoletta, L. Arnaudon, J. Baillie, M. Calviani, F. Caspers, and L. Joergensen, Upgrades and Consolidation of the CERN AD for Operation during the Next Decades in Proceedings of the 4th International Particle Accelerator Conference, IPAC-2013, Shanghai, China, 2013 (Shanghai, China, 2013).

[9] M. Calviani and E. Nowak, CERN Technical Report No. ATS/Note/2012/069(TECH), 2012.

[10] M. D. Church and J. P. Marriner, The Antiproton Sources: Design and Operation, Annu. Rev. Nucl. Part. Sci. 43, 253 (1993).

[11] B. C. Brown, P. Adamson, D. Capista, W. Chou, I. Kourbanis, D. K. Morris, K. Seiya, G. H. Wu, and M. J. Yang, Fermilab main injector: High intensity operation and beam loss control, Phys. Rev. ST Accel. Beams 16, 071001 (2013).

[12] J. Morgan, Fermilab Technical Report No. Pbar Note 683, 2003.

[13] T. W. Eaton, CERN Technical Report No. PS/AA/Note 86-3, 1986.

[14] C. D. Johnson, Present and Future Possibilities of Antiproton Production from Fixed Targets at CERN, Hyperfine Interact. 44, 21 (1989).

[15] N. Simos, H. Ludewig, and N. Catalan-Lasheras, Thermo-mechanical Response of the Halo Intercepts with the SNS Proton Beam in Proceedings of the Particle Accelerator Conference, Chicago, IL, 2001 (IEEE, New york, 2001).

[16] W. Stein, A. Sunwoo, J. C. Sheppard, V. Bharadwaj, and D. C. Schultz, SLAC Technical Report No. SLAC-LCC-0088, 2002.

[17] O. Caretta, T. Davenne, C. Densham, M. Fitton, P. Loveridge, J. ODell, N. Charitonidis, I. Efthymiopoulos, A. Fabich, and L. Rivkin, Response of a tungsten powder target to an incident high energy proton beam, Phys. Rev. ST Accel. Beams 17, 101005 (2014).

[18] A. Ahmad, C. Booth, D. Jenkins, and T. R. Edgecock, Generic study on the design and operation of high power targets, Phys. Rev. ST Accel. Beams 17, 024701 (2014).

[19] A. Dallocchio, Study of Thermo-mechanical effects induced in Solids by High Energy Particle Beams: Analytical and Numerical Methods, Ph.D. thesis, Politecnico di Torino, Torino, 2008.

[20] J. Zukas, Introduction to Hydrocodes (Elsevier Science, New York, 2004).

[21] T. W. Eaton, CERN Technical Report No. PS/AA/Note 84-9, 1984.

[22] D. C. Wilson, C. A. Wingate, J. C. Goldstein, R. P. Godwin, and N. V. Mokhov, Hydrodynamic Calculations of 20-TeV Beam Interactions with the SSC Beam Dump in Particle Accelerator Conference (PAC 93), Washington, DC, USA, 1993 (IEEE, 1993).

[23] B. P. Howell, C. J. Hayhurst, and G. E. Fairlie, AUTODYN Study of the Structural Behaviour of the Main LHC Absorber Block in the Case of a Total Beam Dilution Failure (Century Dynamics Ltd for CERN, Horsham, 2001).

[24] H. Richter, Simulating Transient Effects of Pulsed Beams on Beam Intercepting Devices, Ph.D. thesis, Technischen Universitet Wien, Vienna, 2011.073402-11